# On the effect of non local impact ionization on avalanche multiplication for uniform electric fields.

J. S. Marsland

Department of Electrical Engineering and Electronics

University of Liverpool

Brownlow Hill

Liverpool        L69 3GJ

United Kingdom

Telephone: (44) 151 794 4536

Fax: (44) 151 794 4540

Email: marsland@liv.ac.uk

Abstract

An alternative definition for the non local impact ionization coefficient is given and its relationship to the ionization pathlength probability distribution function (PDF) is determined. A model for the ionization pathlength PDF is proposed and the resultant non local ionization coefficient is derived. An expression for the avalanche multiplication is derived. Results are calculated using non local ionization coefficients for two electric field values using fits to ionization pathlength PDFs calculated using Monte Carlo techniques. These results show two features: the well known dead space effect and at the higher field value a resonance effect. The resonance effect gives rise to level sections in the multiplication curve for multiplications of 2, 4, 8 and 16 if the ionization coefficient ratio is small.

1. Introduction

Avalanche multiplication occurs when energetic carriers create additional carriers by impact ionization. Avalanche photodiodes can amplify signals by avalanche multiplication but this is a random process and it introduces avalanche noise which degrades the signal to noise ratio. The mean square noise current per unit bandwidth, $\langle i^2 \rangle$, is given by the following equation:

$$\langle i^2 \rangle = 2qI_p M^2 F(M) \tag{1}$$

where $q$ is the electronic charge, $I_p$ is the primary photocurrent, $M$ is the multiplication and $F(M)$ is the excess noise factor that takes into account the additional noise produced by the avalanche process. McIntyre [1] derived an expression for $F(M)$ and demonstrated that avalanche noise can be reduced if the electron and hole ionization coefficients are as disparate as possible. McIntyre [1] made the assumption that the ionization coefficients were functions of the local electric field only and he recognised that this restricted his theory to photodiodes with multiplication regions much longer than the 'dead space' length. The dead space is the distance required for a carrier to gain sufficient kinetic energy from the electric field so that it can initiate impact ionization. Although Okuto and Crowell [2] formulated a method for calculating the avalanche multiplication taking into account the dead space there was no similar method for avalanche noise for many years. Consequently much experimental research was directed at finding materials with disparate ionization coefficients and in general the multiplication regions used were long enough to neglect any effect of the dead space.

Van Vliet et al [3] made an early attempt at quantifying the effect of the dead space on noise by assuming that impact ionization could only occur at a finite number of points, each an integral number of dead spaces away from where the initial carrier was generated.  They [3] showed that McIntyre's theory [1] overestimates the excess noise factor when the dead space is a significant proportion of the multiplication region.  However experimentalists continued to view the dead space as only a nuisance to be corrected for e.g. Bulman et al [4] who proposed a correction for the dead space of primary carriers that were injected into the multiplication region.

A number of theoretical papers appeared in the early 1990's.  First Marsland [5] proposed a method for calculating multiplication and excess noise factor that took into account the dead space of both primary carriers and secondary carriers that were created by impact ionization.  However a carrier that has just initiated impact ionization will also have a dead space and this was neglected in the method proposed by Marsland [5].  Next Saleh et al [6] proposed a more complete method that was restricted to only one carrier type initiating impact ionization.  Then Hayat et al [7] extended that work to consider impact ionization initiated by both carrier types. Marsland et al [8] used a hybrid lucky drift / Monte Carlo simulation to estimate the effect of dead space on the excess noise factor.  Finally Hayat et al [9] extended their earlier work to non uniform electric fields.  All of these papers came to the same conclusion: that the excess noise factor could be significantly reduced by the dead space effect as well as by disparate ionization coefficients.

Hu et al [10] made the first systematic experimental study that demonstrated the decrease in excess noise factor as the multiplication width becomes comparable with the dead space length.  Further experimental work [11 - 16] has confirmed this finding.  There is now considerable interest in non local impact ionization leading to new theoretical work [17 - 20] including a new paper by McIntyre [21] and various Monte Carlo studies [22 - 25].

This paper takes a fresh look at non local impact ionization and starts at the most fundamental level by considering the definition of the non local ionization coefficient in Section 2. A model for the ionization pathlength probability distribution function (PDF) is introduced in Section 3 and the resulting non local ionization coefficient is derived. In Section 4 non local ionization coefficients are calculated for two different field values and the results show both the dead space and a resonant behaviour. Section 5 considers avalanche multiplication calculated from non local ionization coefficients. The excess noise factor is considered in the companion paper. Sections 6 and 7 present discussion and conclusions.

2. Definition of the non local impact ionization coefficient

A major problem in the study of non local impact ionization relates to the definition of a non local ionization coefficient. As noted by McIntyre [21] the local ionization coefficient can be unambiguously defined in a number of ways whereas the non local ionization coefficient can be defined in a number of non equivalent ways. The definition used by McIntyre [21] is the same as that used by Okuto and Crowell [2] and the symbol $\alpha_{OC}(z)$ will be used for this definition throughout this paper. The non local ionization coefficient, $\alpha_{OC}(z)$, is defined such that $\alpha_{OC}(z)dz$ is the probability that a carrier will impact ionize in the interval *(z, z + dz)* starting with no kinetic energy at *z = 0* "assuming that it survives to do so without having a previous ionizing collision" in the words of McIntyre [21] or equivalently for a carrier that "has not yet produced an electron hole pair" according to Okuto and Crowell [2]. Using this definition the non local ionization coefficient can be related to the probability distribution function (PDF) for the pathlength to impact ionization. The ionization pathlength PDF, *h(z)*, can be unambiguously defined such that *h(z)dz* is the probability that a carrier undergoes impact ionization for the first time in the interval *(z, z + dz)* starting with no kinetic energy at *z = 0*. The probability, $P_S(z)$, that a carrier travels to *z* without impact ionizing (starting with no kinetic energy at *z = 0*) has been called the survival probability by McIntyre [21] and it is related to the ionization pathlength PDF by the following equation.

$$P_S(z) = 1 - \int_0^z h(x)dx = \int_z^\infty h(x)dx \quad (2)$$

The probability, *h(z)dz*, of a carrier impact ionizing for the first time in the interval *(z, z + dz)* is equal to the survival probability multiplied by $\alpha_{OC}(z)dz$ so that the non local ionization coefficient is given as follows:

$$\alpha_{OC}(z) = \frac{h(z)}{P_S(z)} = \frac{h(z)}{\int_z^\infty h(x)dx} \quad (3)$$

This definition of the non local ionization coefficient only accounts for the first impact ionization suffered by a carrier whereas the following alternative definition accounts for the first and all subsequent ionizations. The non local ionization coefficient, $\alpha(z)$, is defined such that $\alpha(z)dz$ is the probability that a carrier will impact ionize in the interval *(z, z + dz)* starting with no kinetic energy at *z = 0*. The carrier can ionize any number of times in travelling to *z*. Again this definition can be related to the ionization pathlength PDF as follows:

$$\alpha(z) = h(z) + \int_0^z \alpha(x) h(z-x) dx \tag{4}$$

The first term on the right hand side accounts for the first impact ionization and the integral accounts for all other ionizations where a carrier had previously ionized at $z = x$. A detailed justification of this expression is included in the Appendix. An important assumption in the derivation of equation (4) is that the ionization pathlength PDF is the same for subsequent ionizations as it is for the first ionization. This is a commonly made assumption [2, 7, 21] and it requires that a carrier has zero kinetic energy immediately after it initiates impact ionization. If the carrier energy, before ionization, is greater than the band gap then, after ionization, the excess energy must be distributed between the initiating carrier and the two secondary carriers [26]. Taking into account the excess energy using the methodology described in this paper is not straightforward and has not been attempted.

3. Model for the ionization pathlength PDF

A. General case

Spinelli et al [22], Ong et al [25] and Jacob et al [27] have calculated the ionization pathlength PDF, $h(z)$, using Monte Carlo techniques. The characteristic shape of $h(z)$ is a fast rising exponential starting immediately after the dead space followed by a slower decaying exponential. This behaviour can be described by the following expression where $l$ is the length of the dead space region and $a$ and $b$ are constants governing the slope of the rise and fall of $h(z)$.

$$h(z) = \frac{ab}{b-a}\{\exp(-a(z-l)) - \exp(-b(z-l))\}U(z-l) \tag{5}$$

$U(z - l)$ is a unit step function equal to zero for $z < l$ and one elsewhere. The multiplying factor, $ab/(b - a)$, is chosen so that the integral of $h(z)$ from 0 to $\infty$ is unity as appropriate for a PDF. Given a functional form for $h(z)$ it is possible to derive an equation for $\alpha(z)$. First note that the integral in equation (4) is a convolution so that taking the Laplace transform of that expression gives the following:

$$A(s) = H(s) + A(s)H(s) \tag{6}$$

where $A(s)$ is the Laplace transform of $\alpha(z)$ and $H(s)$ is the Laplace transform of $h(z)$ given by the following equation presuming the model proposed above.

$$H(s) = \frac{abe^{-sl}}{(s+a)(s+b)} \tag{7}$$

Now an expression for $A(s)$ can be found by rearranging equation (6) and expressing it in the form of a geometric progression thus:

$$A(s) = \frac{H(s)}{1-H(s)} = \sum_{n=1}^{\infty}\{H(s)\}^n = \sum_{n=1}^{\infty}\frac{a^n b^n e^{-nsl}}{(s+a)^n(s+b)^n} \tag{8}$$

Before the inverse Laplace transform can be applied the expression for $A(s)$ must be rearranged using the following partial fraction formula.

$$\frac{1}{(s+a)^n(s+b)^n} = \sum_{m=1}^{n} \frac{(-1)^{n-m}(2n-m-1)!}{(n-1)!(n-m)!} \left\{ \frac{(b-a)^{m-2n}}{(s+a)^m} + \frac{(a-b)^{m-2n}}{(s+b)^m} \right\} \quad (9)$$

So by substituting equation (9) into equation (8) and performing an inverse Laplace transform the following expression for the non local ionization coefficient can be found:

$$\alpha(z) = \sum_{n=1}^{\infty} h_n(z) \quad (10a)$$

where

$$h_n(z) = a^n b^n U(z-nl) \sum_{m=1}^{n} \frac{(-1)^{n-m}(2n-m-1)!(z-nl)^{m-1}}{(n-1)!(n-m)!(m-1)!} \left\{ \frac{\exp(-a(z-nl))}{(b-a)^{2n-m}} + \frac{\exp(-b(z-nl))}{(a-b)^{2n-m}} \right\}$$

(10b)

The non local ionization coefficient, $\alpha(z)$, is the sum of an infinite number of components, $h_n(z)$, that represent the contribution to $\alpha(z)$ from the $n^{th}$ impact ionization suffered by the carrier. Each $h_n(z)$ is zero for $z < nl$; that is it is offset from the origin by at least a distance $nl$. Therefore only $h_1(z)$ is nonzero for $z < 2l$ so that $\alpha(z) = h_1(z)$ in that interval. Note that $h_1(z)$ is simply $h(z)$ as given by equation (5). The non local ionization coefficient has the following asymptotic value as $z$ tends to infinity.

$$\alpha(z \to \infty) = \left( \frac{1}{a} + \frac{1}{b} + l \right)^{-1} \quad (11)$$

Also of interest is the peak value of $h(z)$ that occurs when $z = z_{max}$.

$$z_{max} = l + \frac{\ln(a/b)}{a-b} \qquad (12)$$

$$h(z_{max}) = \left(\frac{a^b}{b^a}\right)^{\frac{1}{b-a}} \qquad (13)$$

Finally the Okuto and Crowell [2] non local ionization coefficient is given by:

$$\alpha_{OC}(z) = \frac{ab\{1 - \exp((a-b)(z-l))\}}{b - a\exp((a-b)(z-l))} U(z-l) \qquad (14)$$

The asymptotic value of $\alpha_{OC}(z)$ as $z$ tends to infinity is either $a$ or $b$, whichever is smaller.

## B. Special case 1: a = b, 'degenerate'

A degenerate case exists when $a$ and $b$ are equal and in this situation the equations derived above become unusable. The alternative expressions are as follows: first for the ionization pathlength PDF.

$$h(z) = a^2(z-l)\exp(-a(z-l))U(z-l) \qquad (15)$$

This gives rise to the following expression for the non local ionization coefficient using the same procedure as described above.

$$\alpha(z) = \sum_{n=1}^{\infty} \frac{a^{2n}(z-nl)^{2n-1}\exp(-a(z-nl))U(z-nl)}{(2n-1)!} \qquad (16)$$

The asymptotic value of the non local ionization coefficient is:

$$\alpha(z \to \infty) = \left(\frac{2}{a} + l\right)^{-1} \qquad (17)$$

The peak value of $h(z)$ occurs at $z = z_{max}$ where:

$$z_{max} = l + \frac{1}{a} \qquad (18)$$

$$h(z_{max}) = ae^{-1} \qquad (19)$$

Finally the Okuto and Crowell [2] non local ionization coefficient is:

$$\alpha_{OC}(z) = \frac{a^2(z-l)}{a(z-l)+1} U(z-l) \qquad (20)$$

The asymptotic value of $\alpha_{OC}(z)$ as $z$ tends to infinity is $a$.

## C. Special case 2: b → ∞, 'dead space model'

The dead space model has been used as a major simplification in many previous investigations of non local impact ionization including Okuto and Crowell [2], McIntyre [21], Hayat et al [7] and numerous others. In this model the non local ionization coefficient as defined by Okuto and Crowell [2] is assumed to be zero in the dead space region and a constant value thereafter.

$$\alpha_{OC}(z) = aU(z-l) \tag{21}$$

Confusingly (and at the time unknowingly) Marsland [5] applied this simplification to the alternative definition, $\alpha(z)$, rather than $\alpha_{OC}(z)$. The dead space model of Okuto and Crowell [2] and others arises if the coefficient $b$ is infinite. This assumption gives another set of alternative expressions as follows:

$$h(z) = a\exp(-a(z-l))U(z-l) \tag{22}$$

$$\alpha(z) = \sum_{n=1}^{\infty} \frac{a^n(z-nl)^{n-1}\exp(-a(z-nl))U(z-nl)}{(n-1)!} \tag{23}$$

$$\alpha(z \to \infty) = \left(\frac{1}{a}+l\right)^{-1} \tag{24}$$

The peak value of $h(z)$ occurs at $z = z_{max} = l$ and $h(z_{max}) = a$ so that $h(z_{max})$ is always greater than $\alpha(z \to \infty)$ for any non zero value of dead space.

## 4. Non local ionization coefficients derived from Monte Carlo results

Equation (5) can be fitted to the ionization pathlength PDF calculated using Monte Carlo techniques [27] for electrons in GaAs at a field of $3 \times 10^7$ Vm$^{-1}$. The fit is shown in Figure 1 and the values of the parameters used are $a = 1$ μm$^{-1}$, $b = 10$ μm$^{-1}$ and $l = 0.15$ μm. Figure 2 shows the non local ionization coefficients for this parameter set. The solid bold line is $\alpha(z)$ as calculated by equation (10) whereas the solid fine line is $\alpha_{OC}(z)$ as calculated using equation (14). The dotted lines show $h_n(z)$ for $n = 1$ to 7. For $0 < z < 0.3$ μm, that is $0 < z < 2l$, $\alpha(z)$ follows $h(z)$ displaying both the dead space and the fast exponential rise. After $h(z)$ peaks $\alpha(z)$ continues to rise and reaches a plateau at $z = 0.5$ μm well before the peak in $h_2(z)$ at $z = 1.5$ μm. The non local ionization coefficient then remains constant even though the $h_n(z)$ components are constantly changing. The distance between successive $h_n(z)$ peaks is equal to the reciprocal of the steady state ionization coefficient; roughly equal to 9 dead spaces.

Jacob et al [27] have also calculated the ionization pathlength PDF for electrons in GaAs at a higher field of $10^8$ Vm$^{-1}$. Again this result can be fitted using the following parameters: $a = b = 300$ μm$^{-1}$ and $l = 0.0415$ μm. The non local ionization coefficients for these parameter values are shown in Figure 3, the bold line for $\alpha(z)$ and the fine line for $\alpha_{OC}(z)$. A remarkably different behaviour is observed compared to the result for the lower field strength. The ionization coefficient, $\alpha(z)$, now consists of a series of isolated peaks, each peak lower and broader than the previous one. The n$^{th}$ peak is due to $h_n(z)$; that is the n$^{th}$ impact ionization and the peaks begin to merge for $n > 4$. The peak value of $h(z)$ is more than 5 times greater than $\alpha(z \to \infty)$ and the distance between successive $h_n(z)$ peaks is roughly equal to the reciprocal of $\alpha(z \to \infty)$; approximately 1.2 dead spaces.

Transient features in the non local ionization coefficient have been observed before using Monte Carlo techniques. Kim and Hess [28] and Plimmer et al [17] both found transients in the electron ionization coefficient in GaAs at a field of approximately $7 \times 10^7$ Vm$^{-1}$. The oscillations shown in [17] appear much larger than those shown in [28] although this may simply be because $\alpha(z)$ is plotted on a linear scale in [17] and a logarithmic scale in [28]. This work offers a simple explanation of these transient effects; if $h(z_{max})$ is greater than $\alpha(z\rightarrow\infty)$ then an overshoot must occur. If an overshoot does occur and $h(2l)$ is less than $\alpha(z\rightarrow\infty)$ then oscillations will occur. If $h(z)$ has decayed to near zero for $z < 2l$ then severe oscillations are inevitable as shown in Figure 3. The period of the oscillations (and more generally the spacing of $h_n(z)$) is related to the reciprocal of the ionization coefficient, $\alpha(z\rightarrow\infty)$, rather than the dead space length.

## 5. Mean avalanche multiplication

### A. Theory

Consider a multiplication region of width $W$ occupying the interval $0 < z < W$. Electrons are assumed to travel under the influence of an electric field in the negative $z$ direction and holes in the positive $z$ direction. This is the convention used by McIntyre [1, 21] and Marsland [5] and is the opposite of that used by Okuto and Crowell [2] and Hayat et al [7]. Now define $M(z)$ as the mean multiplication for an electron hole (e-h) pair created at a point $z$ in the multiplication region. The electron can impact ionize and create further e-h pairs in the interval $0 < x < z$ and the hole can create e-h pairs in the interval $z < x < W$. A secondary e-h pair created at $x$ will also have a mean multiplication given by $M(x)$. The probability that an electron starting at $z$ with no kinetic energy will impact ionize in the interval $(x, x + dx)$ is $\alpha(z - x)dx$. Similarly define a non local ionization coefficient for holes, $\beta(x - z)$, such that $\beta(x - z)dx$ is the probability that a hole starting at $z$ with no kinetic energy will impact ionize in the interval $(x, x + dx)$. The mean multiplication at $z$, $M(z)$, is the sum of the original e-h pair and other e-h pairs created by the original electron or hole at $x$ multiplied by *M(x)* and weighted by the appropriate probability. This gives the following equation for the mean multiplication.

$$M(z) = 1 + \int_0^z \alpha(z-x)M(x)dx + \int_z^W \beta(x-z)M(x)dx \qquad (25)$$

This equation is a generalization of equation (1) of McIntyre's 1966 paper [1] where the electron and hole ionization coefficients were assumed to be functions of electric field only. Marsland [5] also used equation (25) but assumed a uniform field and a dead space model for $\alpha(z)$ and $\beta(z)$. Although the derivation of equation (25) is explained above in terms of e-h pairs created in the body of the multiplication region it also applies to injected carriers i.e. electrons created at $z = W$ or holes created at $z = 0$. The equation is also applicable to nonuniform fields if the ionization coefficients are considered to be functions of both *x* and *z* rather than *(x - z)* although it is only applied to uniform fields in the remainder of this paper.

## B. Results

The multiplication, *M(z)*, has been calculated using the non local electron ionization coefficient, $\alpha(z)$, shown in Figure 2 and for a number of different values of the ionization coefficient ratio, *k*. For this paper the ionization coefficient ratio, *k*, is defined as the hole ionization coefficient divided by the electron ionization coefficient as *z* tends to infinity.

$$k = \frac{\beta(z \to \infty)}{\alpha(z \to \infty)} \qquad (26)$$

The parameters used for $\alpha(z)$ are $a = 1$ μm$^{-1}$, $b = 10$ μm$^{-1}$ and $l = 0.15$ μm and the same parameters are used for $\beta(z)$ when $k = 1$. For $k = 0.3$, 0.1 and 0.03 there are many ways of choosing the parameters for $\beta(z)$ in order to satisfy equation (26). In this work the dead space length for holes is assumed to be equal to that for electrons for all values of $k$. Two criteria for selecting the parameters $a$ and $b$ for holes have been considered. Firstly the ratio of $a$ to $b$ can be the same as that for electrons and secondly the larger parameter can be held constant ($b = 10$ μm$^{-1}$) and the smaller parameter ($a$) change. The resulting non local ionization coefficients are shown in Figure 4. The dashed lines (constant ratio) approach the asymptotic value more slowly than the solid lines (constant $b$). Using the constant ratio scheme for $k = 0.03$ the hole ionization coefficient does not reach its asymptotic value in the first 4 μm, approximately the width of the multiplication region required for avalanche breakdown. Therefore in the following results the $k$ value is selected by changing the $a$ parameter whilst keeping $b$ and $l$ constant.

Figure 5 shows $M(z)$ for $W = 2$ μm and for $k = 0.3$ (top), 0.1, 0.03 and 0 (bottom). The results were calculated using the meshing (400 mesh points) and iteration method used by Marsland [5], Hayat et al [7] and McIntyre [21]. The number of iterations required for convergence increases with $k$ and the maximum multiplication and is less than 40 for the results in Figure 5. No results are shown for $k = 1$ because avalanche breakdown occurs for $W > 1.5$ μm using the parameters assumed here. The multiplication is 1 for $k = 0$ if the e-h pair is created within one dead space of the left hand boundary. This is expected because neither holes ($\beta = 0$) nor electrons can ionize. To the right of this region the multiplication increases as $z$ increases because electrons have a greater distance in which to ionize. As $k$ (and $\beta$) increases the multiplication increases at all points due to the additional gain provided by hole initiated ionization. There is a clear minimum in the multiplication for $k = 0.3$ at approx. $z = 0.15$ μm. Only holes created in the interval $0 < z < 0.15$ μm can contribute to multiplication because of the dead space for electrons. However holes created closer to $z = 0$ have a greater distance in which to ionize and therefore the multiplication increases slightly as $z$ approaches 0. In the middle of the multiplication region multiplication increases with $z$ because the electron ionization coefficient is greater than the hole ionization coefficient and the interval in which the initial electron can ionize increases with increasing $z$.

The multiplication, $M(W)$, for electrons injected at $z = W$ is plotted against multiplication width $W$ in Figure 6. The solid lines show the multiplication calculated using non local ionization coefficients whereas the dashed lines are calculated using local ionization coefficients with $\alpha = 0.8$ μm$^{-1}$ for all values of $z$. In each case the five curves taken from left to right relate to $k = 1, 0.3, 0.1, 0.03$ and 0. The results demonstrate that the inclusion of the non local effects gives lower multiplication for the same multiplication region width (which is equivalent to voltage for a fixed electric field as assumed here). Avalanche breakdown occurs in a wider multiplication region (higher voltage) if non local effects are taken into account.

The difference between local and non local calculations can be fully explained in terms of a dead space model. The non local results shown in Figure 6 can be almost exactly reproduced if a dead space model is applied to $\alpha(z)$ and $\beta(z)$; that is $\alpha(z) = 0$ and $\beta(z) = 0$ for $z < D$ and $\alpha(z) = 0.8$ μm$^{-1}$ and $\beta(z) = k\alpha(z)$ for $z > D$. However the distance, $D$, is not equal to the dead space, $l$, which is 0.15 μm for both carriers in these calculations. Rather $D$ is equal to 0.221 μm; that is an effective dead space that takes into account the transition from $\alpha(z) = 0$ to $\alpha(z \rightarrow \infty)$. The effective dead space,

*D*, is given by the solution of equation (27) for $z \to \infty$. That is the area under the non local ionization coefficient as shown in Figure 2 is equal to the area under a 'dead space model' ionization coefficient that is zero for $z < D$ and $\alpha(z \to \infty)$ for $z > D$.

$$(z - D)\alpha(z) = \int_0^z \alpha(x)dx \tag{27}$$

Now consider multiplication at the higher electric field of $10^8$ Vm$^{-1}$ where the non local ionization coefficient displays overshoot and oscillatory behaviour as shown in Figure 3. The electron ionization coefficient is given by $a = b = 300$ µm$^{-1}$ and $l = 0.0415$ µm. Figure 7 shows two possible candidates for the hole ionization coefficient for $k = 0.3$ (top), $k = 0.1$ (middle) and $k = 0.03$ (bottom). The solid lines are calculated using $b = 300$ µm$^{-1}$ and $l = 0.0415$ µm with *a* selected to produce the required *k*. The dashed lines are calculated using $a = b$ and $l = 0.0415$ µm with *a* and *b* selected for *k* and again the ionization coefficient approaches the asymptotic value more slowly if calculated with the ratio of *a* to *b* fixed. The results that follow have been calculated with *b* and *l* constant and *k* selected by changing *a*.

Figure 8 shows *M(z)* for $W = 0.12$ µm and for $k = 0.3$ (top), 0.1, 0.03 and 0 (bottom). The multiplication is 1 for $k = 0$ and $0 < z < 0.0415$ µm because the dead space of an electron created in this interval reaches to the end of the multiplication region. The multiplication is 2 for $k = 0$ and $0.06$ µm $< z < 0.083$ µm because an electron created in this interval will ionize once and only once before reaching the end of the multiplication region. The secondary electron created by this ionization can not initiate ionization due to the dead space. An electron injected into the multiplication region at $z = W = 0.12$ µm has a multiplication of 4 for $k = 0$. The injected electron will ionize twice (and only twice), once in the middle of the multiplication region and once towards the end. The secondary electron created in the middle of the device will ionize once resulting in three new electrons created in total so that $M(W) = 4$. As $k$ (and $\beta$) increases the multiplication increases because hole initiated ionization provides extra gain. There are now distinct maxima in the multiplication for $k = 0.3$. Consider the maximum in $M(z)$ at approximately $z = 0.105$ μm where the initial hole will not ionize due to its dead space. If the initial electron is created at a $z$ less than that of the maximum then the probability that it will ionize twice before reaching $z = 0$ decreases. Therefore it will create fewer secondary holes (on the average) and there will be less multiplication overall. If the initial electron is created at a $z$ greater than that of the maximum then it will ionize at higher values of $z$ (on the average) and the secondary holes created will have a smaller region in which to ionize. Therefore there will be less hole initiated ionization and, as a result, the multiplication will decrease. Similar considerations apply to the maximum in $M(z)$ at approximately $z = 0.055$ μm.

Figure 9 shows the multiplication, $M(W)$, for injected electrons against multiplication region width, $W$, for $k = 1, 0.3, 0.1, 0.03$ and $0$. The dashed lines show the multiplication calculated using local ionization coefficients with $\alpha = 20.76$ μm$^{-1}$ and the solid lines show the multiplication calculated using non local ionization coefficients. For both sets of results $k$ increases from right to left. Again these results calculated for the higher field value ($10^8$ Vm$^{-1}$) are distinctly different from the lower field results shown in Figure 6. The inclusion of non local effects gives lower multiplication for the same multiplication region width. However the multiplication does not increase as a smooth curve as before but has level sections at $M(W) = 2, 4, 8$ and $16$. Furthermore the results can not be explained in terms of the dead space alone.

If the multiplication is calculated using the dead space model for $\alpha(z)$ and $\beta(z)$ with $D = 0.0239$ μm then the shift to greater widths is roughly reproduced but not with the level sections. This value of the effective dead space width, $D$, is calculated for electrons using equation (27) and is not necessarily appropriate for holes (except for $k = 1$ and $k = 0$). Notably $D$ is now less than the actual dead space width, $l$, and multiplication calculated using the dead space model with the actual dead space (i.e. $D = l$) greatly overestimates the shift to greater multiplication widths. The level sections occur because the avalanche is more deterministic in these regions. Injected electrons will ionize a fixed number of times equal to the number of peaks in $\alpha(z)$ traversed (as shown in Figure 3). The secondary electrons will also ionize a given number of times and hole initiated ionization will be negligible.

## C. Comparison with previous work

There are obvious similarities between this work and that of Marsland [5]. The main difference is not in the derivation of equation (25) but in the understanding and use of the non local ionization coefficient. Marsland [5] offered no concise definition of the non local ionization coefficient but did recognise that the 'dead space' of a carrier immediately after that carrier had initiated impact ionization was not accounted for by the use of a dead space model for $\alpha(z)$ and $\beta(z)$. The alternative definition for $\alpha(z)$ given in equation (4) does account for this 'dead space' whereas $\alpha_{OC}(z)$ given in equation (3) does not. Therefore the full picture presented here requires the use of equations (25) and (4) together. Furthermore the dead space model approximation for $\alpha(z)$ as used by Marsland [5] is unphysical because it is impossible to find a $h(z)$ that will give $\alpha(z) = 0$ for $z < D$ and a constant $\alpha(z)$ for $z \geq D$ for any non zero value of $D$.

There are similarities also between this work and that of Hayat et al [7] who avoided the definition of $\alpha(z)$ by only using $h(z)$. Their approach leads to two interlinked equations that can be iterated in the same way as equation (25). Equations (10) and (11) of Hayat et al [7] can be shown to be mathematically equivalent to equation (25) of this paper for the special case of $k = 0$ and give numerically identical results for any value of $k$. The claim made by Hayat et al [7] that Marsland [5] substantially underestimated the effect of the dead space because the dead space of secondary carriers was neglected is false. On the contrary Marsland [5] (and this work) does account for the dead space of the secondary carriers created by impact ionization. The explanation of the difference between Hayat et al [7] and Marsland [5] is actually more mundane. The results presented in both papers assume a fixed multiplication width, $W$, and a maximum multiplication (for no dead space) of 40 in [7] and 2.5 in [5]. Figures 6 and 9 (max. mult. = 20) show that the effect of the dead space increases greatly as breakdown is approached ($M \to \infty$). This is the only substantive difference in the multiplication results reported by [7] and [5].

## 6. Discussion

The results show that non local impact ionization gives rise to two different behaviours; the well known dead space effect and a resonance effect that has not previously been reported. The dead space effect is important when the dead space width (that is a distance a carrier requires to gain sufficient energy for impact ionization) is a significant fraction of the multiplication width. The resonance effect is important when the dead space width is comparable to the reciprocal of the ionization coefficient, $1/\alpha(z\to\infty)$. The dead space effect gives a decrease in the avalanche multiplication. When the resonance effect is also present the avalanche multiplication is not only decreased but also displays level sections where it changes little with increasing multiplication width (or voltage).  These level sections occur at multiplications of 2, 4, 8 and 16 for $k = 0$ and tend to disappear for higher values of the ionization coefficient ratio, $k$.

No observations of these resonant effects have been made in avalanche photodiodes and such a practical demonstration may be technologically unfeasible.  Firstly it is generally held that the ionization coefficient ratio tends to 1 at very high fields, e.g. as shown in [4] although the concept of 'bulk' ionization coefficients is not very useful at these very high fields.  Secondly the breakdown voltage of photodiodes exhibiting resonant effects will be comparable to the breakdown voltage of a tunnel diode so that the multiplication current may be masked by a tunnelling current.  The breakdown voltages implied by Figures 3 and 6 may be a little higher than expected because the ionization pathlength PDF was fitted to Monte Carlo simulations [27] that assumed a high threshold voltage of 4.1 eV.

The analysis of both non local effects is greatly advanced using the methods described in this paper.  For instance Monte Carlo techniques may need several hundred or more electron flights to derive the non local ionization coefficient in a simulated device at one electric field value.  The same computer power could be utilised to find the ionization pathlength PDF at many field values using Monte Carlo.  Then the non local ionization coefficient can be calculated using the method described in sections 3 and 4 at those field values and at intermediary values by interpolation of the fitting parameters $a$, $b$ and $l$.  The non local ionization coefficient can then be used to calculate the avalanche multiplication in ultra thin avalanche photodiodes using the method described in section 5.

## 7. Conclusion

An alternative definition for the non local ionization coefficient, $\alpha(z)$, has been proposed and its relationship to the ionization pathlength PDF has been determined. A model for the ionization pathlength PDF has been proposed and the resulting $\alpha(z)$ derived. The 'dead space model' [2, 7, 21] is a special case of the proposed model. The model has been fitted to ionization pathlength PDFs calculated by Monte Carlo techniques for two electric field values and oscillations in $\alpha(z)$ are observed at the higher field value. Avalanche multiplication has been calculated using the non local ionization coefficient and, in addition to the well known dead space effect which is observed at both field values, a 'resonance' effect is observed at the higher field value. The resonance effect gives rise to level sections in the multiplication curve at $M = 2$, 4, 8, and 16 for a small ionization coefficient ratio.

## Acknowledgements

I thank Professor R. C. Woods for a critical reading of this manuscript.

Appendix

The ionization pathlength PDF, *h(z)*, is defined in section 2 such that *h(z)dz* is the probability that a carrier undergoes impact ionization for the first time in the interval (*z, z + dz*) starting with no kinetic energy at *z* = 0. Now define $h_2(z)dz$ as the probability that a carrier undergoes impact ionization for the second time in the interval (*z, z + dz*) starting with no kinetic energy at *z* = 0. Assuming that the carrier returns to zero kinetic energy after the first ionization then the probability that the carrier ionizes for the second time at *z* is *h(z-x)dz* given the first ionization at *z = x*. The probability of a first ionization in the interval (*x, x + dx*) followed by a second ionization in the interval (*z, z + dz*) is

$$h(x)dx.h(z-x)dz \qquad (A1)$$

and integrating this over all *x* in the range 0 to *z* gives $h_2(z)dz$

$$h_2(z) = \int_0^z h(x)h(z-x)dx \qquad (A2)$$

The same argument applies for $h_3(z)dz$ : the probability that a carrier undergoes ionization for the third time in the interval (*z, z + dz*) starting with no kinetic energy at *z* = 0.

$$h_3(z) = \int_0^z h_2(x)h(z-x)dx \qquad (A3)$$

and in general

$$h_n(z) = \int_0^z h_{n-1}(x)h(z-x)dx \qquad n > 1 \qquad (A4)$$

where $h_1(z) = h(z)$. The non local ionization coefficient is the sum of all these separate probabilities up to some number, $N$, determined by the multiplication region width and the field strength.

$$\alpha(z) = \sum_{n=1}^{N} h_n(z) \qquad (A5)$$

Now substituting the equation (A4) into equation (A5) gives:

$$\alpha(z) = h(z) + \sum_{n=2}^{N} \int_0^z h_{n-1}(x)h(z-x)dx \qquad (A6)$$

Rearranging gives:

$$\alpha(z) = h(z) + \int_0^z \sum_{n=1}^{N-1} h_n(x)h(z-x)dx \qquad (A7)$$

The upper limit of summation has changed from $N$ to $(N - 1)$ but $N$ is not a constant - just a sufficiently large number for a given device - so the summation can be replaced by $\alpha(x)$ giving equation (4) of section 2.

References.

1   R. J. McIntyre, "Multiplication noise in uniform avalanche diodes," *IEEE Trans. Electron Devices*, vol. ED-13, pp.164-168, Jan. 1966.

2   Y. Okuto and C. R. Crowell, "Ionization coefficients in semiconductors: A nonlocalized property," *Phys. Rev. B*, vol. 10, pp. 4284-4296, Nov. 1974.

3   K. M. van Vliet, A. Friedmann, and L. M. Rucker, "Theory of carrier multiplication and noise in avalanche devices – Part II: Two carrier processes," *IEEE Trans. Electron Devices*, vol. ED-26, pp.752-764, May 1979.

4   G. E. Bulman, V. M. Robbins, and G. E. Stillman, "The determination of impact ionization coefficients in (100) gallium arsenide using avalanche noise and photocurrent multiplication measurements," *IEEE Trans. Electron Devices*, vol. ED-32, pp.2454-2466, Nov. 1985.

5   J. S. Marsland, "On the effect of ionization dead spaces on avalanche multiplication and noise for uniform electric fields," *J. Appl. Phys*., vol. 67, pp.1929-1933, Feb. 1990.

6   B. E. A. Saleh, M. M. Hayat, and M. C. Teich, "Effect of dead space on the excess noise factor and time response of avalanche photodiodes," *IEEE Trans. Electron Devices*, vol. ED-37, pp.1976-1984, Sept. 1990.

7   M. M. Hayat, B. E. A. Saleh, and M. C. Teich, "Effect of dead space on gain and noise of double carrier multiplication avalanche photodiodes," *IEEE Trans. Electron Devices*, vol. ED-39, pp. 546-552, Mar. 1992.

8   J. S. Marsland, R. C. Woods, and C. A. Brownhill, "Lucky drift estimation of excess noise factor for conventional avalanche photodiodes including the dead space effect," *IEEE Trans. Electron Devices*, vol. ED-39, pp. 1129-1135, May 1992.

9   M. M. Hayat, W. L. Sargeant, and B. E. A. Saleh, "Effect of dead space on gain and noise in Si and GaAs photodiodes," *IEEE J. Quantum Electron.*, vol. 28, pp. 1360-1365, May 1992.

10  C. Hu, K. A. Anselm, B. G. Streetman, and J. C. Campbell, "Noise characteristics of thin multiplication region GaAs avalanche photodiodes," *Appl. Phys. Lett.*, vol. 69, pp. 3734-3736, Dec. 1996.

11  K. F. Li, D. S. Ong, J. P. R. David, R. C. Tozer, G. J. Rees, P. N. Robson, and R. Grey, "Low noise GaAs and $Al_{0.3}Ga_{0.7}As$ avalanche photodetectors," *IEE Proc. Optoelectron.*, vol. 146, pp.21–24, Feb. 1999.

12  P. Yuan, K. A. Anselm, C. Hu, H. Nie, C. Lenox, A. L. Holmes, B. G. Streetman, J. C. Campbell, and R. J. McIntyre, "A new look at impact ionization – Part II: Gain and noise in short avalanche photodiodes," *IEEE Trans. Electron Devices*, vol. ED-46, pp.1632-1639, Aug. 1999.

13  P. Yuan, C. C. Hansing, K. A. Anselm, C. V. Lenox, H. Nie, A. L. Holmes, B. G. Streetman, and J. C. Campbell, "Impact ionization charateristics of III-V semiconductors for a wide range of multiplication thicknesses," *IEEE J. Quantum Electron.*, vol. 36, pp.198-204, Feb. 2000.

14  K. F. Li, D. S. Ong, J. P. R. David, R. C. Tozer, G. J. Rees, S. A. Plimmer, K. Y. Chang and J. S. Roberts, "Avalanche noise characteristics of thin GaAs structures with distributed carrier generation," *IEEE Trans. Electron Devices*, vol. ED-47, pp.910-914, May 2000.

15  C. H. Tan, J. C. Clark, J. P. R. David, G. J. Rees, S. A. Plimmer, R. C. Tozer, D. C. Herbert, D. J. Robbins, W. Y. Leong, and J. Newey, "Avalanche noise measurement in thin Si $p^+$-$i$-$n^+$ diodes," *Appl. Phys. Lett.*, vol. 76, pp.3926-3928, June 2000.

16  C. H. Tan, J. P. R. David, S. A. Plimmer, G. J. Rees, R. C. Tozer, and R. Grey, "Low multiplication noise thin $Al_{0.6}Ga_{0.4}As$ avalanche photodiodes,", *IEEE Trans. Electron Devices*, vol. ED-48, pp.1310 - 1317, July 2001.

17  S.A. Plimmer, J.P.R. David, and G.M. Dunn, "Spatial limitations to the application of the lucky drift theory of impact ionization," *IEEE Trans. Electron Devices*, vol. ED-44, pp.659-663, Apr. 1997.

18  D. C. Herbert, "Avalanche noise in submicrometre *pin* diodes," *Electron. Lett.*, vol. 33, pp. 1257-1258, July 1997.

19  M. A. Saleh, M. M. Hayat, B. E. A. Saleh, and M. C. Teich, "Dead space based theory correctly predicts excess noise factor for thin GaAs and AlGaAs avalanche photodiodes," *IEEE Trans. Electron Devices*, vol. ED-47, pp. 625-633, Mar. 2000.

20  C. H. Tan, J. P. R. David, G. J. Rees, R. C. Tozer, and D. C. Herbert, "Treatment of soft threshold in impact ionization," *J. Appl. Phys.*, vol. 90, pp. 2538-2543, Sept. 2001.

21  R. J. McIntyre, "A new look at impact ionization - Part I: A theory of gain, noise, breakdown probability and frequency response," *IEEE Trans. Electron Devices*, vol. ED-46, pp.1623-1631, Aug. 1999.

22  A. Spinelli, A. Pacelli, and A. L. Lacaita, "Dead space approximation for impact ionization in silicon," *Appl. Phys. Lett.*, vol. 69, pp. 3707 - 3709, Dec. 1996.

23  S. A. Plimmer, J. P. R. David, and D. S. Ong, "The merits and limitations of local impact ionization theory," *IEEE Trans. Electron Devices*, vol. ED-47, pp. 1080-1088, May 2000.

24  S. A. Plimmer, J. P. R. David, R. Grey, and G. J. Rees, "Avalanche multiplication in $Al_xGa_{1-x}As$ ($x$ = 0 to 0.6)," *IEEE Trans. Electron Devices*, vol. ED-47, pp. 1089-1097, May 2000.

25  D. S. Ong, K. F. Li, S. A. Plimmer, G. J. Rees, J. P. R. David, and P. N. Robson, "Full band Monte Carlo modeling of impact ionization, avalanche multiplication, and noise in submicron GaAs $p^+$-$i$-$n^+$ diodes", *J. Appl. Phys.*, vol. 87, pp.7885-7891, June 2000.

26  E. O. Kane, "Electron scattering by pair production in silicon," *Phys. Rev.*, vol. 159, pp. 624-631, July 1967.

27  B. Jacob, S. A. Plimmer, P. N. Robson, and G. J. Rees, "Lucky drift model for nonlocal impact ionisation," *IEE Proc.-Optoelectron.*, vol. 148, pp. 81-83, Feb. 2001.

28  K. Kim and K. Hess, "Simulations of electron impact ionization rate in GaAs in nonuniform electric fields," *J. Appl. Phys.*, vol. 60, pp. 2626-2629, Oct. 1986.

Figure Captions.

Fig. 1　Equation 5 fitted to $h(z)$ derived using Monte Carlo techniques [27] for electrons in GaAs at an electric field of $3 \times 10^7$ Vm$^{-1}$

Fig. 2　Non local ionization coefficients $\alpha(z)$ (solid bold line), $\alpha_{OC}(z)$ (solid fine line) and $h_n(z)$ (dotted lines) for an electric field of $3 \times 10^7$ Vm$^{-1}$

Fig. 3　Non local ionization coefficients $\alpha(z)$ (solid bold line) and $\alpha_{OC}(z)$ (solid fine line) for an electric field of $10^8$ Vm$^{-1}$

Fig. 4　Non local ionization coefficients calculated using equation (10) with $l = 0.15$ μm and $b = 10$ μm$^{-1}$ (solid lines) or $a = b/10$ (dashed lines) with $a$ or $a$ and $b$ selected to give $\alpha(z \to \infty) = 0.8$ μm$^{-1}$ (top), 0.24 μm$^{-1}$, 0.08 μm$^{-1}$ or 0.024 μm$^{-1}$ (bottom).

Fig. 5　Multiplication, $M(z)$, calculated for a 2 μm wide multiplication region using non local ionization coefficients for an electric field of $3 \times 10^7$ Vm$^{-1}$ with $k = 0.3$ (top), 0.1, 0.03 and 0 (bottom).

Fig. 6　Multiplication for injected electrons, $M(W)$, against multiplication region width, $W$, using non local ionization coefficients (solid lines) and local ionization coefficients (dashed lines) for $k = 1, 0.3, 0.1, 0.03$ and 0 with $k$ increasing from right to left for both sets of curves.

Fig. 7　Non local ionization coefficients calculated using equation (10) with $l = 0.0415$ μm and $b = 300$ μm$^{-1}$ (solid lines) or $a = b$ (dashed lines) with $a$ or $a$ and $b$ selected to give $\alpha(z \to \infty) = 6.228$ μm$^{-1}$ (top), 2.076 μm$^{-1}$ (middle) or 0.6228 μm$^{-1}$ (bottom).

Fig. 8   Multiplication, *M(z)*, calculated for a 0.12 μm wide multiplication region using non local ionization coefficients for an electric field of $10^8$ Vm$^{-1}$ with *k* = 0.3 (top), 0.1, 0.03 and 0 (bottom).

Fig. 9   Multiplication for injected electrons, *M(W)*, against multiplication region width, *W*, using non local ionization coefficients (solid lines) and local ionization coefficients (dashed lines) for *k* = 1, 0.3, 0.1, 0.03 and 0 with *k* increasing from right to left for both sets of curves.

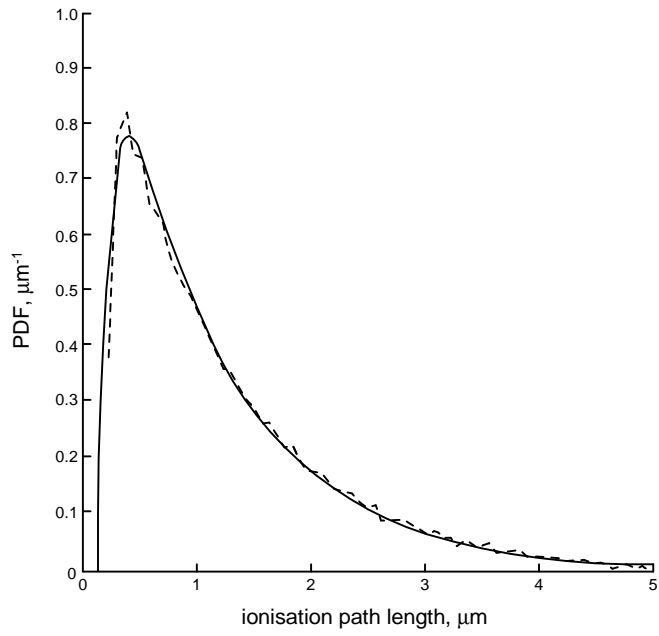

Figure 1.

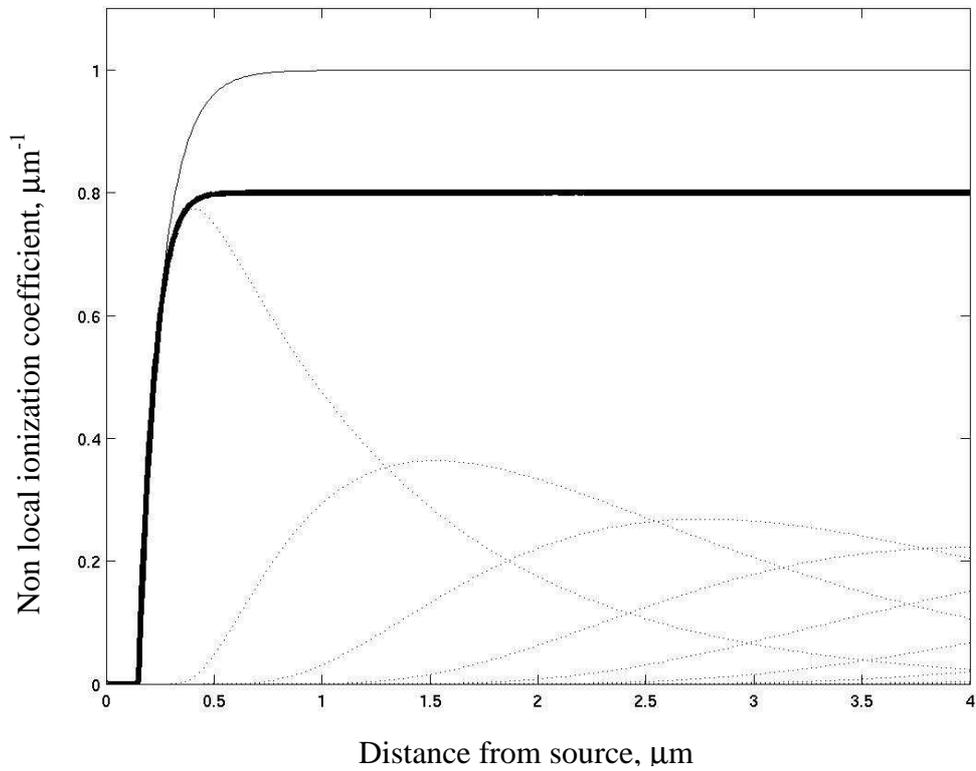

Distance from source, µm
Figure 2.

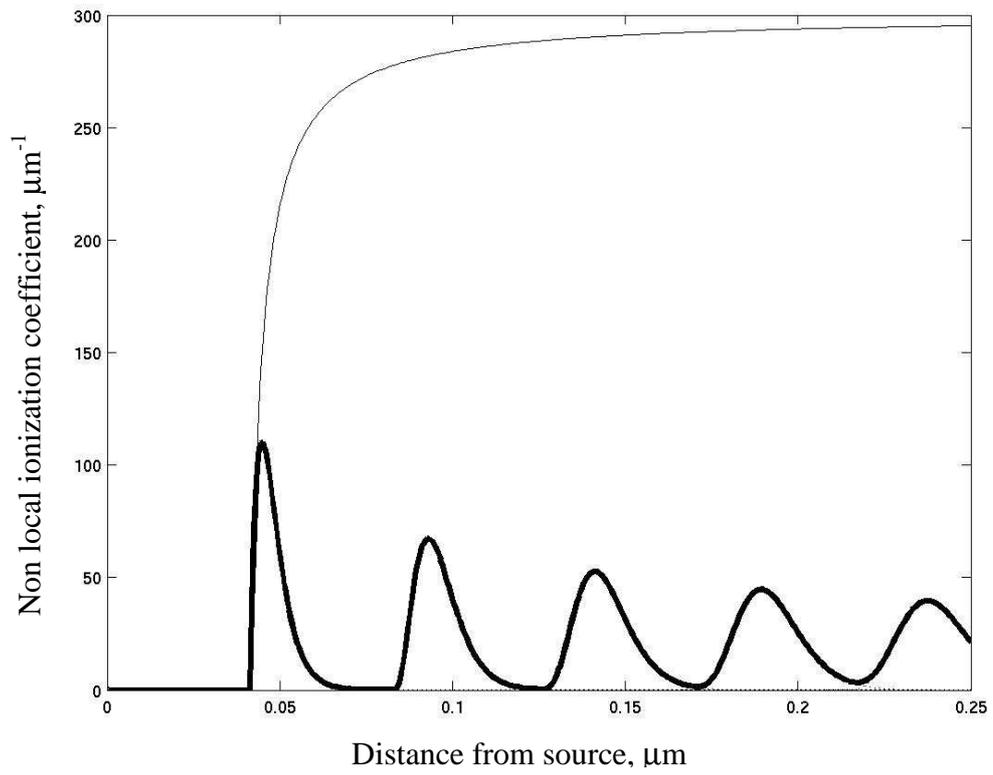

Figure 3.

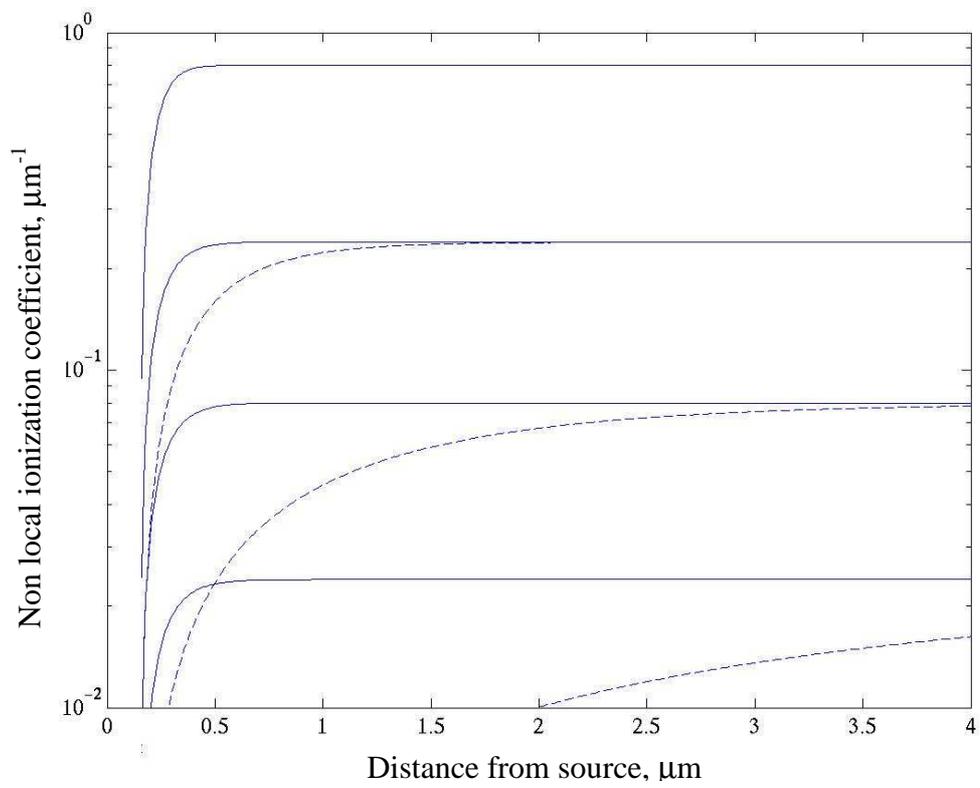

Figure 4.

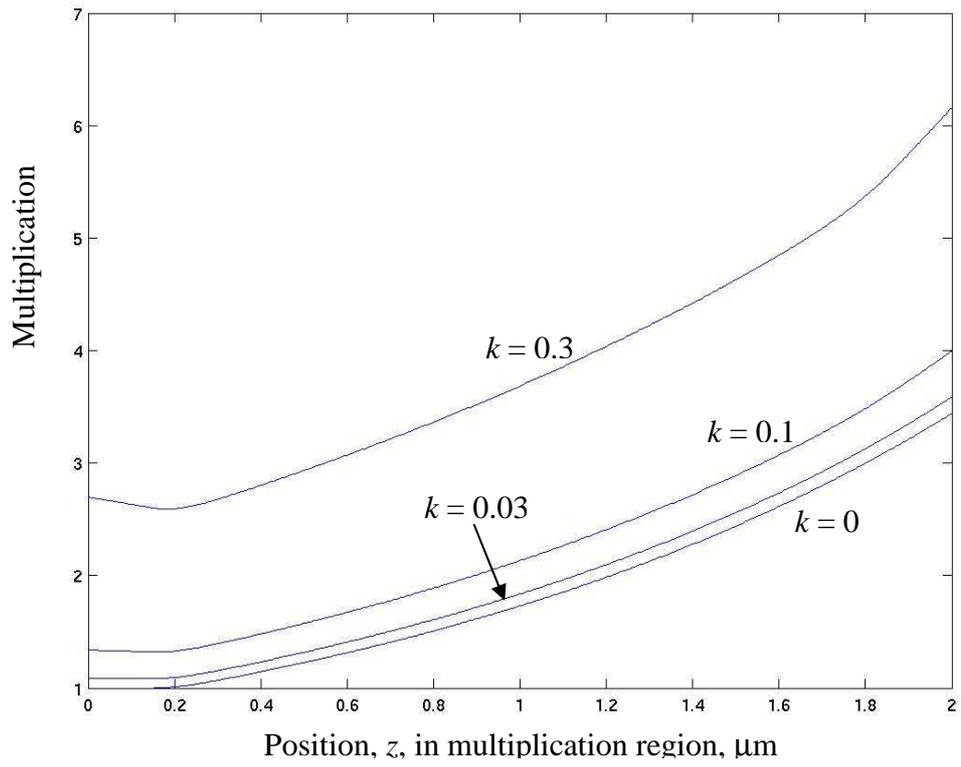

Figure 5.

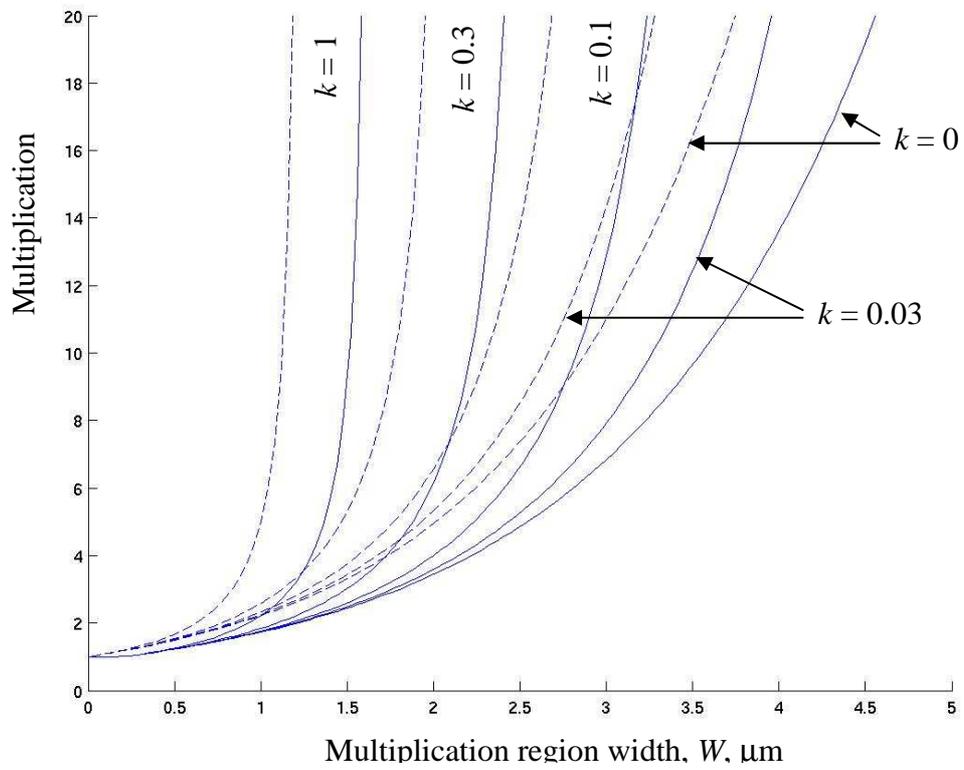

Figure 6.

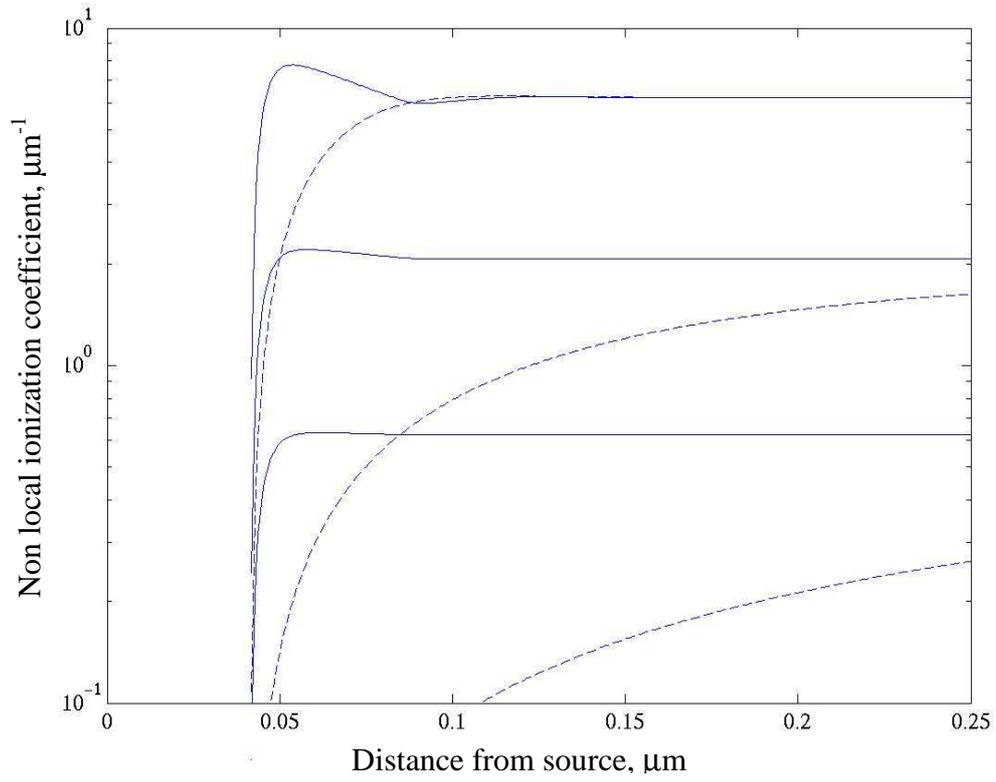

Figure 7.

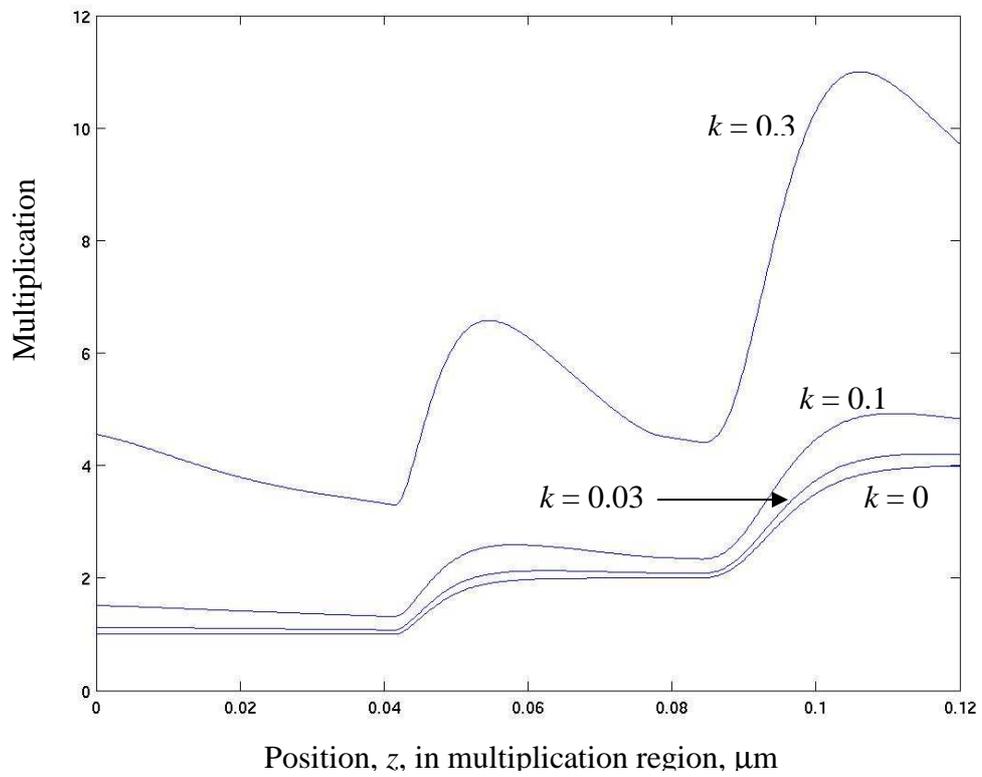

Figure 8.

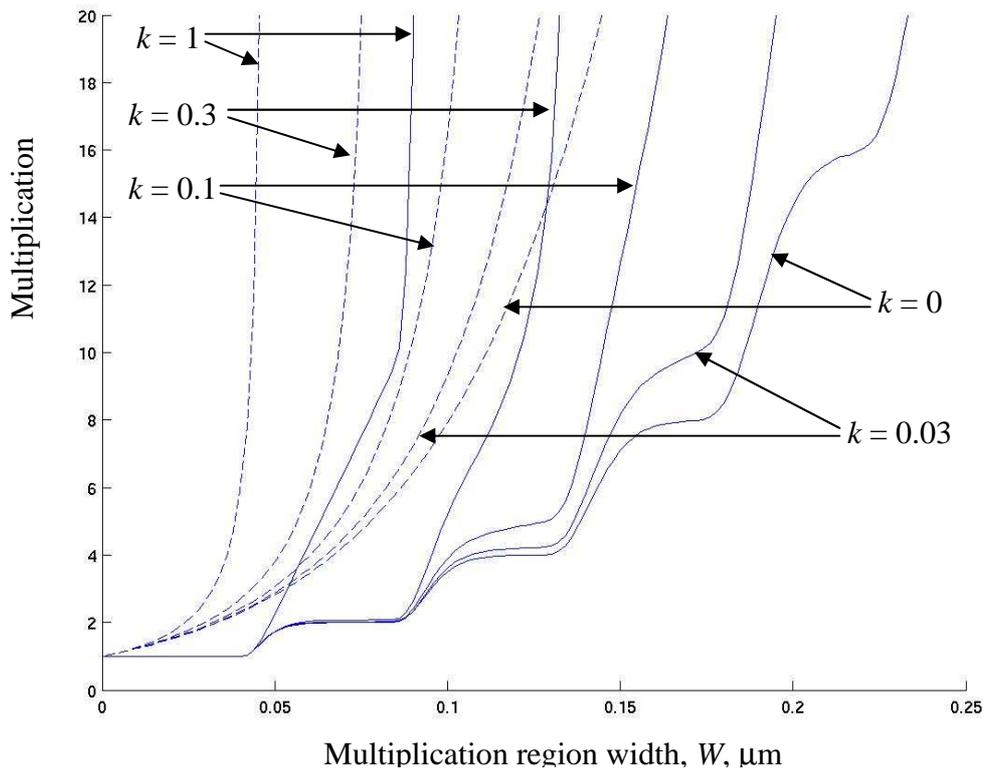

Figure 9